# Coexistence of in-plane and out-of-plane exchange Bias in correlated kagome antiferromagnet Mn$_{3-x}$Cr$_x$Sn

Xiaoyan Yang[1], Yi Qiu[2], Jie Su[2], Yalei Huang[1], Jincang Zhang[1, 3, †], Xiao Liang[4, ‡], Guixin Cao[1, ⋆]

[1] Materials Genome Institute, Shanghai University, 200444 Shanghai, China

[2] College of Chemistry and Molecular Engineering, Peking University, Beijing 100871, China

[3] Zhejiang Laboratory, Hangzhou 311100, China

[4] Information Materials and Device Applications Key Laboratory of Sichuan Provincil Universities, Chengdu University of Information Technology, 610225 Sichuan, China

**Abstract:**

The materials exhibiting exchange bias (EB) have been extensively investigated mainly due to their great technological applications in magnetic sensors, but its underlying mechanism remains elusive. Here we report the novel coexistence of in-plane and out-of-plane EB in the Cr-doped Mn$_3$Sn, a non-colinear antiferromagnet with a geometrically frustrated Kagome plane of Mn. Field-cooling experiments with the applied field parallel and perpendicular to the frustrated Kagome plane exhibits loop shifts and enhanced coercivities. Interestingly, a maximum EB field of 1090 Oe is observed along out-of-plane direction in the Mn$_{2.58}$Cr$_{0.42}$Sn sample, higher than that of in-plane value. Our results indicate that the exchange bias along perpendicular kagome plane is primarily induced by Dzyaloshinskii-Moriya interactions due to the breaking of interfacial symmetry, while the EB along kagome plane is due to the exchange effects at interface of AFM and FM component originating from the net moment. These findings provide a new insight on EB in the kagome AFM materials, which are important and highly potential to the application of antiferromagnetic spintronics.

**Introduction**

The exchange bias effect usually results from the interfacial exchange coupling between antiferromagnetic (AFM) and ferromagnetic (FM) part, resulting in unidirectional anisotropy [1-3]. The EB phenomenon is reflected in the magnetic hysteresis loop by a shift along the magnetic field axis [4], which arises after the sample is cooled through the Néel point in a magnetic field. Based on the traditional EB model assumed uncompensated spins [5], the AFM spins remains in the state they were frozen into during the process, while the FM is rotated and are responsible for the spin memory. In a system with a compensated interface [3], the EB models based on domain state in dilated AFMs[4] and symmetry properties [6] have been proposed. Despite the extensive theoretically study, understanding the full nature of EB effect remains a great challenge[7, 8].

⋆ Corresponding author, G. Cao, Email: guixincao@shu.edu.cn
‡ Corresponding author, X. Liang, E-mail: liangxiao920@163.com
† Corresponding author, J. Zhang, E-mail: jczhang@shu.edu.cn

Numerous experimental investigations on exchange biased FM/AFM bilayers and multilayers have been performed [9-15] to understand the mechanism behind and to develop its potential applications in spintronic devices. The magnetic anisotropy of AFM and the ratio of AFM and FM determine the exchange bias [16, 17]: large AFM anisotropy benefits the shift of hysteresis loop, while the AFM and FM spins may rotate simultaneously for a small AFM anisotropy, inducing the coercivity only. In this Letter, we explore the Exchange bias and micromagnetic configurations in both parallelly and perpendicularly magnetized single crystals of Cr doped $Mn_3Sn$. Undoped $Mn_3Sn$ is a hexagonal Weyl semimetal, in which noncolinear antiferromagnetic configuration is in the $x$-$y$ plane with a stacked kagome lattice [18]. The geometrical frustration leads to a 120° structure of Mn moments, whose symmetry allows a very small spin canting and thus producing spontaneous weak ferromagnetism (approximately 0.002 $\mu_B$/Mn) tilted toward the easy axis $x$ ($2\bar{1}\bar{1}0$) [19]. With such a vanishingly small anisotropy energy, the coupling of the weak magnetization to the magnetic field plays an essential role for the magnetic field control of the sublattice moments [20-22]. Therefore, it provides a great condition for the generation and study of EB in the correlated AFM system. In fact, recent work have observed the EB phenomena in $Mn_3Sn$ [22] and similar AFM thin films [3], and alloys[16]. Here we show that Cr doping induces additional FM component to the in-plane direction and causes a symmetry broken in the out-of-plane direction. As a result, large EB effects occur in both in-plane and out-of-plane direction, which origin from the enhanced exchange coupling of AFM/FM interface and Dzyaloshinskii-Moriya (DM) interaction induced by the broken interface inversion symmetry, respectively. Our $Mn_{3-x}Cr_xSn$ series enable us to tune the $H_{EB}$ by changing the composition $x$, thereby outlining a route towards the design of organic spintronic devices.

**Experimental method**

$Mn_{3-x}Cr_xSn$ single crystals were grown by the Sn-flux growth method [23, 24]. The crystals were characterized by energy-dispersive x-ray spectroscopy (EDX; HGST FlexSEM-1000) and scanning electron microscopy (SEM). To obtain more detailed structural information, single crystal XRD data of $Mn_{3-x}Cr_xSn$ was collected at 300 K on a Rigaku XtaLAB Synergy diffractometer, with Mo Kα radiation ($\lambda$ = 0.71073 Å). Data reduction and empirical absorption correction were performed using the CrysAlisPro program. Magnetic property measurements were performed in a Quantum Design Magnetic Property Measurement System (MPMS). First-principles calculations based on the density functional theory (DFT) are performed by VASP5.4. Hexagonal $Mn_3Sn$ adopts a $D0_{19}$-type structure with space group $P6_3/mmc$ at 300 K according to our single crystal XRD results (Fig.1S), consistent with previous report [20]. With Cr content until $x$=0.42, no obvious structural transition occurs (Fig.1S(c) and (d)). According to the symmetry of hexagonal lattice, we mark it in three directions $x$ [$2\bar{1}\bar{1}0$], $y$ [$01\bar{1}0$]) and $z$ [0001]. In Mn$_3$Sn, Sn lies in the center of the Mn hexagonal ring forming a Kagome lattice (Fig. 1S($a$)). With increasing Cr until $x$-0.42, the lattice constants $a$ and $c$ decrease with no structural transition observed as displayed in Fig.1($c$) and ($d$).

## Results and discussion

We first present in Fig.1(a) and (b) the calculated magnetic structure for Mn$_{3-x}$Cr$_x$Sn ($x$=0, 0.125). In pure Mn$_3$Sn, the in-plane Mn moments shows a 120° spin structure with a uniform negative vector chirality, so called inverse triangular antiferromagnetic structure. The magnetic moment per Mn atom is about 3.167 $\mu B$, and the in-plane spin direction is a little canted from 120° due to DM interactions, which leads to a weak net ferromagnetic moment of 0.002 $\mu B$/Mn, consistent with the previous reports[20]. The calculated magnetic moment of Cr atom is 3.04 $\mu B$, which is a little lower than Mn atoms, and the spin directions of Mn nearby the Cr atoms tilted both in in-plane and out-plane directions as shown in Fig.1(b). Comparing with the spin distributions in pure Mn$_3$Sn (see Fig.1(a)), the in-plane spin directions of upper layer Mn (Mn2, Mn5) and lower layer Mn (Mn3, Mn6) are left-handed and right-handed rotated 2~3 degrees, respectively. However, the in-plane spin directions of Mn4 and Cr1 little rotated (~±0.3 degree). Furthermore, the spin direction of all atoms trend to tilted from in-plane to out-plane directions. Correspondingly, we show that the magnetic texture sensitively determines the vary of the band structure as shown in Fig.1(c) and (d). The pure Mn$_3$Sn holds two pairs of Weyl points near the Fermi-level (-0.1~0.1eV), and one of them is overlaped due to band folding in supper cell. While in Mn$_{2.875}$Cr$_{0.125}$Sn, the nontrivial band features are disappeared, which means topological phase transition caused by spin structure changes. Moreover, we would expect striking exchange bias because of the DM interaction and magnetic exchange effect from the weak ferromagnetic moment and inverse triangular antiferromagnetic structure.

To identify the effects of the Cr-doping induced variations in spin structure, we did the magnetic property measurements on high quality Mn$_{3-x}$Cr$_x$Sn single crystals ($x$ = 0~0.42). Similar to previous observations [24], Susceptibility ($\chi$) of Mn$_3$Sn undergoes an inverse triangular AFM to spiral phase transition at $T_1$ = 270 K, then exhibits a ferromagnetic-like transition at $T_2$ = 200 K as shown in Fig.2(a). Upon further cooling, the zero-field cooling (ZFC) and field fooling (FC) $\chi$ curves shows a clear bifurcation, consistent with the spiral and glass magnetic states coexistence [25]. Note that the magnetic structure at low temperature (low-T) sensitively depends on the synthesis method and element contents [24-28]. With increasing Cr content, both $T_1$ and $T_2$ are obviously suppressed and shift to lower temperature for both $B//x$ and $B//z$ axis in $x$ = 0.06 as shown in Fig.2S. When $x$ = 0.16, the $T_1$ transition is completely suppressed and $T_2$ undergoes further decreasing to lower temperature $T_2$ = 150 K for $B//x$ and $B//z$ when Cr content increases to $x$ = 0.42 as displayed in Fig.2S (e) and (j). Therefore, our results here shows that the spiral structure transition $T_1$ is completely suppressed by Cr content with $x=0.16$, comparable with the previous report that of $x=0.15$ [29]. Here, we argue that compared with pure Mn₃Sn, spiral structure disappears at low temperature in the Cr-doped Mn$_{3-x}$Cr$_x$Sn and produces an inverse triangular spin structure. Meanwhile, the existence of a small irreversibility between ZFC and FC curves suggests that the low temperature phase is not perfectly AFM in nature. We identify the coexistence of AFM and FM phase at low-T region, which provides the exhibition of Cr-doping tuned EB in the system. To verify the possible EB effect, we have measured the *M(H)* under FC 1 T and -1 T respectively, for both *B//*x and *B//*z axis. As shown in Fig.2(c) and (d), we observed the shift of the hysteresis loop in negative (positive) field direction by about 1899 Oe and 2100 Oe when applied 1 T (-1 T) field for *B//*x and *B//*z axis, respectively. It can be noted that the produced value of $H_{EB}$ and coercive field ($H_C$) is nearly the same for the FC 1 T and -1 T directions for both *B//*x and *B//*z. Therefore, we conclude that the observed EB behavior is intrinsic for both in-plane and out-of-plane of Mn$_{2.72}$Cr$_{0.42}$Sn. Here, $H_{EB}$ and $H_C$ are calculated using $H_{EB} = -(H_L + H_R)/2$ and $H_C = -(H_L + H_R)/2$, where $H_L$ and $H_R$ are the left and right coercivity [30].

Mn₃Sn displays weak ferromagnetic behavior due to the magnetic frustration from the instability inverse triangular structure as we calculated. Through Cr-doping, the net moment originating from the inversed non-colinear spin structure varies to *z* axis (Table IS and IIS). By applying a magnetic field, we can manipulate the magnetization direction of Mn₃Sn domains. Hence, to investigate in detail the EB effect with applied field parallel and perpendicular to the kagome plane, respectively, in the Cr-doped Mn₃Sn system, we measured the in-plane (*B//x*) and out-of-plane (*B//z*) magnetization loops using a FC protocol, as shown in Fig. 3. In the field cooling process, a magnetic field of +1 T is applied at 300 K, and the samples are cooled to 2 K and the loops are then recorded. The $+B_{FC}$ loop shifts to a negative field direction, indicating the existence of EB in Mn$_{3-x}$Cr$_x$Sn. $H_{EB}$ and $H_C$ exhibit a monotonic increase upon increasing Cr content. Strikingly, additional Cr doping causes more than a tenfold increment in exchange bias field when compared with that in pure Mn₃Sn along both *x-y* plane and *z* axis. However, the $H_{EB}$ for *B//z* is much larger than that of *B//x* and *B//y*, though the *Hc* along three axis's are comparable with each other (Fig.4(b)). Notably, at 2 K, the Mn₃Sn produces a $H_{EB}$ of 70 Oe in FC mode. While, with $x= 0.42$, the perpendicular exchange bias possesses a maximum value of 1070 Oe (Fig 4(a)). Moreover, the sudden increase of $H_{EB}$ occurs at $x=0.16$, after that, the increasing Cr content cause the change of easy axis from *x-y* plane to *z* axis along with strong uniaxial anisotropy in Mn$_{3-x}$Cr$_x$Sn as shown in Fig.4(a).

To further investigate the physical origin of doping induced $H_{EB}$ enhancement, we calculated the magnetocrystalline anisotropy energy (MAE) in Cr$_x$Mn$_{3-x}$Sn as shown in Fig.4 (c). Here, MAE is defined as the total energy difference of the spin aligns to $[2\bar{1}\bar{1}0]$ and [0001] direction in Mn₃Sn unit cell. In pure Mn₃Sn, MAE is calculated for -1.2 meV per unit cell, that indicates weak in-plane magnetic anisotropy of this material.

With creasing Cr substitution content, MAE increases and changes to positive when substitution content $x$ exceeding 0.125, which indicates the out-of-plane magnetic anisotropy. Our experiments results are consistent with the calculations [Fig. 3 (c)]. The anisotropy transformation may be due to the lattice strain *ie.* Jahn-Teller effect, which would significantly enhance the out-of-plane DMI. Meanwhile, by subtracting the AFM background from the hysteresis loop, we calculated the net magnetic moment of per Mn/Cr atom for $B//x$ and $B//z$ (Fig.4(d)). As expected, the net magnetic moment tends to increase with increasing Cr content. Moreover, its trend with increasing Cr content calculated from experimental results is consistent with that from the theoretical results as shown in Fig. 4 (d).

As we know, traditionally, the unidirectional anisotropy and exchange bias can be qualitatively understood by assuming exchange interaction at the AFM-FM interface [1-3]. In our present system, the unidirectional anisotropy and the magnetic environment for in-plane and out-of-plane are obviously different. Moreover, the EB anisotropy is clear. Therefore, the anisotropy evolution and relevant origin of EB exhibited need to be discussed separately.

***In-plane EB model.*** The EB of $Mn_3Sn$ thin film with field parallel to the kagome plane have been investigated based on the AFM-FM interface model [22]. Through Cr-doping, Cr atom sits on the Mn site, which induce the in-plane geometric frustration and a little rotation of the in-plane spin directions of Mn4 and Cr1 (Fig.1 (c) ). Accordingly, the WFM domain increase due to the double-exchange interaction of $Mn^{2+}$-$Sn^{6-}$-$Cr^{3+}$ [31]. Therefore, the AFM component in undoped part and the increasing WFM due to Cr-doping creates exchange coupling interface, hence holding a mechanism consistent with AFM-FM interface model in nature [32]. The described in-plane model was displayed in Fig. 5(a). The phenomenological formula of the $H_{EB}$ can be expressed as $H_{EB}^{in} = (S_F S_{AF} \Sigma J_{ij} \cos\theta_{ij})/(M_F t_F)$, where $S_F$ and $S_{AF}$ are the spins of the magnetic moments, and the angle between them ($\cos\theta_{ij}$. $J_{ij}$ ) is the spin-spin interaction strength between $S_F$ and $S_{AF}$. $M_F$ and $t_F$ are the magnetization and thickness of the FM domain, respectively [33]. With increasing Cr doping, $S_F$ increases due to the magnetic frustration in the in-plane part. Meanwhile, the value of $J_{ij}$ increase also owing to the additional FM component. Therefore, the significant increased in- plane EB originates from the enhancement of the ferromagnetic spins and interaction strength.

***Out-of-plane EB model.*** With increasing Cr content, we have observed the enhanced MAE as shown in Fig.4, illustrating the significantly enhanced out-of-plane DMI. In addition, taking $Mn_{2.875}Cr_{0.125}Sn$ as an example, the calculated magnetic moment of Cr atom is 3.04 $\mu B$, which is a little lower than Mn atoms, and the spin directions of Mn nearby the Cr atoms tilted both in in-plane and out-plane directions. Therefore, with the increasing Cr content, the out-plane magnetic moments are significantly increased, which indicates DMI increased near Cr doped area of the material, and this effect is damping away from the doping center, predicting 2D-Skyrmions in $Cr_xMn_{3-x}Sn$. All these indicates that the observed large perpendicular EB with applied field perpendicular to the kagome plane originates from the enhanced DMI. The DMI induced perpendicular EB was previously observed in the FM/AFM layer systems with perpendicular magnetic anisotropy [3, 34]. Depending on the calculated spin structure and observed MAE, we give a sketched picture to explain the large out-of-plane EB as shown in Fig. 5(b). $H_i^{dm}$ is the DM field experienced by a Mn or Cr spin, originating from the interaction of atoms between neighboring kagome planes. The DM interactions across the interface induce an effective magnetic field acting on the triangular sublattice $|H_{cell}^{dm}| = |\Sigma_{i=1}^{3} H_i^{dm}|$, which points normal to the interface. An exchange bias arises due to the DM interactions with the EB field can be expressed as $H_{EB}^{dm} = (|H_{cell}^{dm}| \cos \alpha)/(M_F t_F)$, where $\alpha$ is the angle of FM magnetization with respect to the effective DM field [3]. On the one hand, the spin orbit scattering of the embedded Cr ions enhance the strength of the DM interaction [35]. In present system, Cr-doping breaks the inversion symmetry of neighboring layers across the interface, causing a large increase of $H_{cell}^{dm}$. Moreover, the increasing Cr content induces uniaxial anisotropy when $x>0.16$, causing the increase of $\alpha$. We indicate that the contribution of these two parameters results in a large perpendicular EB.

Comparing the two models, we indicate that the in-plane EB is based on an exchange coupling interaction across neighboring AFM/FM domains, while the out-of-plane EB achieves interlayer pinning effect via the DM-interaction. Notably, NN atoms in the same layer also generate nonnegligible DM-interaction, but do not contribute to the EB effect [3]. Shuai Dong et al. reported that the EB will not occur if $\mu_0 H_{FC}$ is perpendicular to DM field in the ideal case [6]. Moreover, N. C. Koon used full micromagnetic calculations to illustrate that the interface exchange coupling is relatively strong when the FM/AFM axis are oriented perpendicularly [36], similar to the classical "spin-flop" state in bulk antiferromagnets [37, 38]. This theory explains the strong vertical coupling behavior in our experiments. Moreover, the out-of-plane blocking temperature is much smaller than that of in-plane temperature, demonstrating that the contribution of the DM-interaction to EB is more likely to disappear with thermally disturbed in this system. It would be interesting to verify this origin in future study.

We will now turn our focus to the details of the temperature and cooling field dependence of EB effect at three axes for $Mn_{2.72}Cr_{0.42}Sn$ with the maxim EB effect. As shown in Fig. 5(c) and (d), with increasing temperature, the offset of loops decreases gradually and disappears completely above blocking temperature. Importantly, the in-plane (300 K) and out-of-plane blocking temperature (50 K) shows a significant difference, which further illustrates the different mechanism of the EB. The hysteresis loops in different direction after various field cooling are shown in Fig. 5(e) and (f). In ZFC mode, we can observe that samples still present $H_{EB}$, showing a value of 112 Oe in $z$ direction [Fig. 4(f)]. The origin of this spontaneous EB is that partial ferromagnetic ordering induces the exchange anisotropy during the virgin magnetization, which is needed for the zero field cooled EB [39]. In the FC mode, the $H_{EB}$ and $H_C$ show a nonmonotonic change, and $H_C$ exhibits the same trend with that of $H_{EB}$. It was reported that the responsible mechanism for $H_C$ enhancement is the domain wall pinning due to defects at the FM/AFM interface [40], which explains the comparable value for both in-plane and out-of-plane in present systems. With cooling field $B_{FC}$ increasing, the out-of-plane $H_{EB}$ increases initially and achieves a maximum value of 1090 Oe at $B_{FC}$ = 0.1 T. However, the maximum in-plane $H_{EB}$ is 1029 Oe at $B_{FC}$ = 0.5 T. The peaks of out-of-plane and in-plane $H_{EB}$ appear in different cooling fields. The EB field is usually thought of as a balance, which is between the Zeeman energy of the FM domain and the exchange coupling energy at the interface [41]. Therefore, the cooling field of $B_{FC}$ = 0.1 T and 0.5 T can be considered as an effective depinning threshold field, at which the Zeeman coupling can overcome the magnetic interaction. The results show that the spin-spin exchange coupling and interlayer DM interaction have different threshold fields. Ultimately, Combining the results of Fig. 4 (c) and (f), The $B_{FC}$ and temperature dependence of EB effect confirms that the EB in different directions is strongly anisotropic.

In summary, we theoretically found that Cr-doping tune the inverse triangular antiferromagnetic structure of $Mn_3Sn$ tilting to out-of-plane, inducing large net moment with enhanced MAE and enhanced DMI along perpendicular to the kagome plane. Correspondingly, the coexistence of in-plane and out-of-plane EB were experimentally observed in the AFM systems. Strikingly, additional Cr doping causes more than a tenfold increment in exchange bias field when compared with that in pure $Mn_3Sn$, which is the origin of the perpendicular EB observed here. The found of the large EB in the typical AFM system with frustrated kagome structure provides a new avenue for the correlation mechanism of the symmetry breaking and DMI.

**Acknowledgements**:

Supported by Key Research Project of Zhejiang Lab (No. 2021PE0AC02) and the National Natural Science Foundation of China (NSFC, Grant No. 51902033), the China Postdoctoral Science Foundation (Grant No. M2020683282), and the Scientific Research Foundation of CUIT (Grant No. KYTZ202006). A portion of this work was performed on the Steady High Magnetic Field Facilities, High Magnetic Field Laboratory, CAS. JS acknowledges the support by the NNSFC (Grant No. 22003003).

**Figure captions:**

**Fig. 1.** The calculated magnetic structures and band structures for $Mn_{3-x}Cr_xSn$ ($x$=0, 0.125). The Angles between the moments and $x$, $y$ and $z$ axis are $\alpha$, $\beta$ and $\gamma$. (a) In pure $Mn_3Sn$, the in-plane Mn moments shows a 120° spin structure with a uniform negative vector chirality. (b) With $x = 0.125$, the in-plane spin directions of upper layer Mn (Mn2, Mn5) and lower layer Mn (Mn3, Mn6) are left-handed and right-handed rotated 2~3 degrees, respectively. The in-plane spin directions of Mn4 and Cr1 little rotated (~±0.3 degree). (c) The band structure of pure $Mn_3Sn$ holds two pairs of Weyl points near the Fermi-level (-0.1~0.1eV), and one of them is overlaped due to band folding in supper cell. (d) The nontrivial band features are disappeared in $Mn_{2.875}Cr_{0.125}Sn$, which means topological phase transition caused by spin structure changes.

**Fig. 2**. With (a) B//x and (b) B//z axis, the temperature dependence of susceptibility ($\chi$) measured in zero field cooled (ZFC, black lines) and field cooled (FC, red lines) modes for $Mn_{3-x}Cr_xSn$. The shift of hysteresis loops at 2 K in negative (positive) field direction when applied 1 T (-1 T) cooling field for (c) B//x and (d) B//z axis, respectively.

**Fig. 3**. M(H) loops of $Mn_{2.72}Cr_{0.42}Sn$ measured with $B_{FC}$ = 1 T at 2 K after FC, field parallels to (a-e) x and (f-j) z direction, respectively. With Cr content increasing, the bias of M(H) loops increases gradually. Meanwhile, the curves display different shapes for in-plane and out-of-plane.

**Fig. 4.** (a) Exchange bias fields $H_{EB}$ and (b) coercive fields $H_C$ for various Cr content x at 2 K. When x > 0.16, inverse triangular spin AFM and WFM phases coexist. (c) The calculated results of magnetocrystalline anisotropy energy in $Mn_{3-x}Cr_xSn$. With increasing x, the easy axis gradually shifts from in-plane to out-of-plane. (d) The net magnetic moment of per Mn or Cr atom dependence with increasing Cr content x. The green pentacles represent the calculation results.

**Fig. 5.** (a) The sketched In-plane model: magnetic exchange coupling at the interface between the AFM and FM domains at $B_{FC}$//x. Green and yellow spheres represent Cr and Sn atoms. The red and blue spheres represent the Mn atoms in the upper and under layers, respectively. (b) Out-of-plane model: Picture of the magnetic order at the (0001) interface in neighboring atomic layers. The interaction of the lower and upper layer is marked with black arrows. A DM interface field $H_i^{dm}$, and a DM field acting on the Cr and Mn atoms per unit cell $H_{cell}^{dm}$ are displayed by purple and red arrows. (c) Exchange bias fields $H_{EB}$ and (d) coercive fields $H_C$ at B//x, B//y and B//z as a function of temperature for $Mn_{2.72}Cr_{0.42}Sn$, respectively. The inset of (c) enlarges the $H_{EB}$ above 50 K. The in-plane and out-of-plane blocking temperatures are 300 K and 50 K, respectively. M(H) loops of $Mn_{2.72}Cr_{0.42}Sn$ measured with different cooling fields (0 ~ 5 T) at T = 2 K for applied field parallel to (d) x and (f) z direction, respectively.

Table. I The calculated results of net magnetic moment in $Mn_{3-x}Cr_xSn$ (x = 0~0.5).

|  | x = 0 | x = 0.125 | x = 0.25 | x = 0.375 | x = 0.5 |
|---|---|---|---|---|---|
| Net magnetic moment ub/Mn(Cr) | 0.002 | 0.016 | 0.024 | 0.029 | 0.032 |

Fig.1

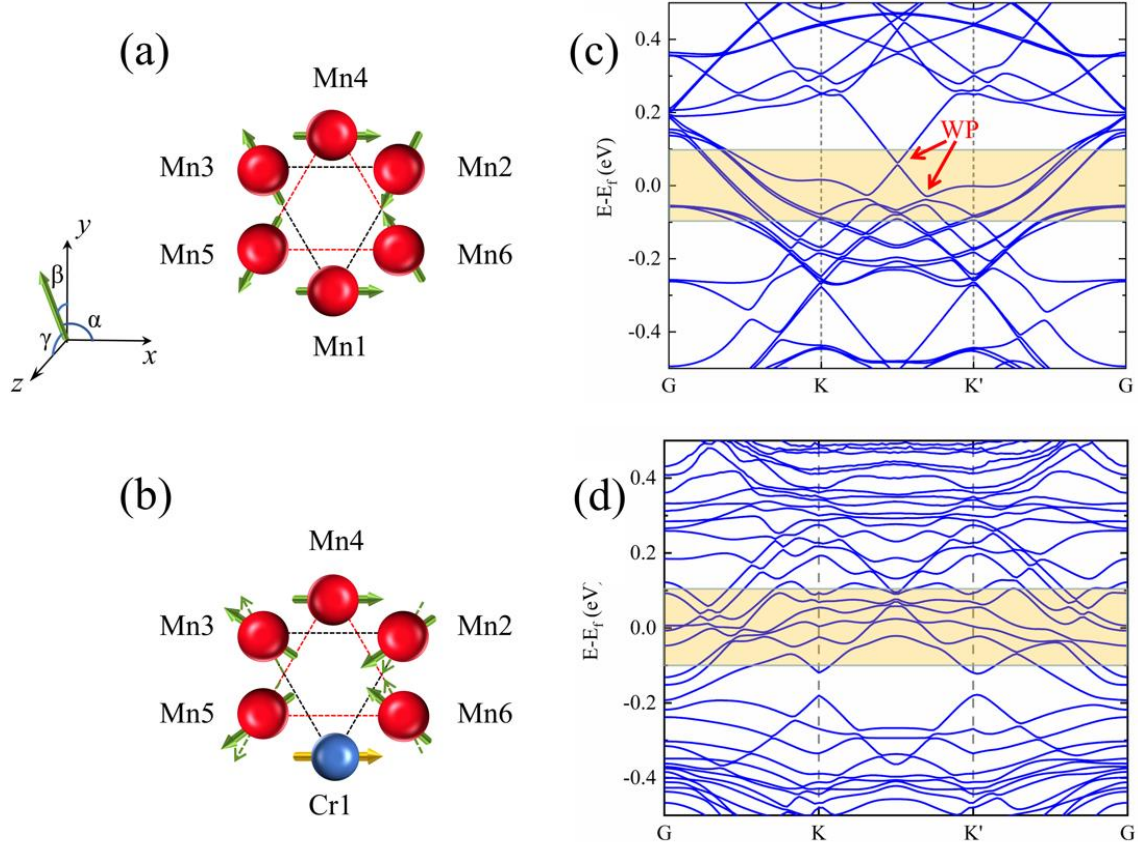

Fig.2.

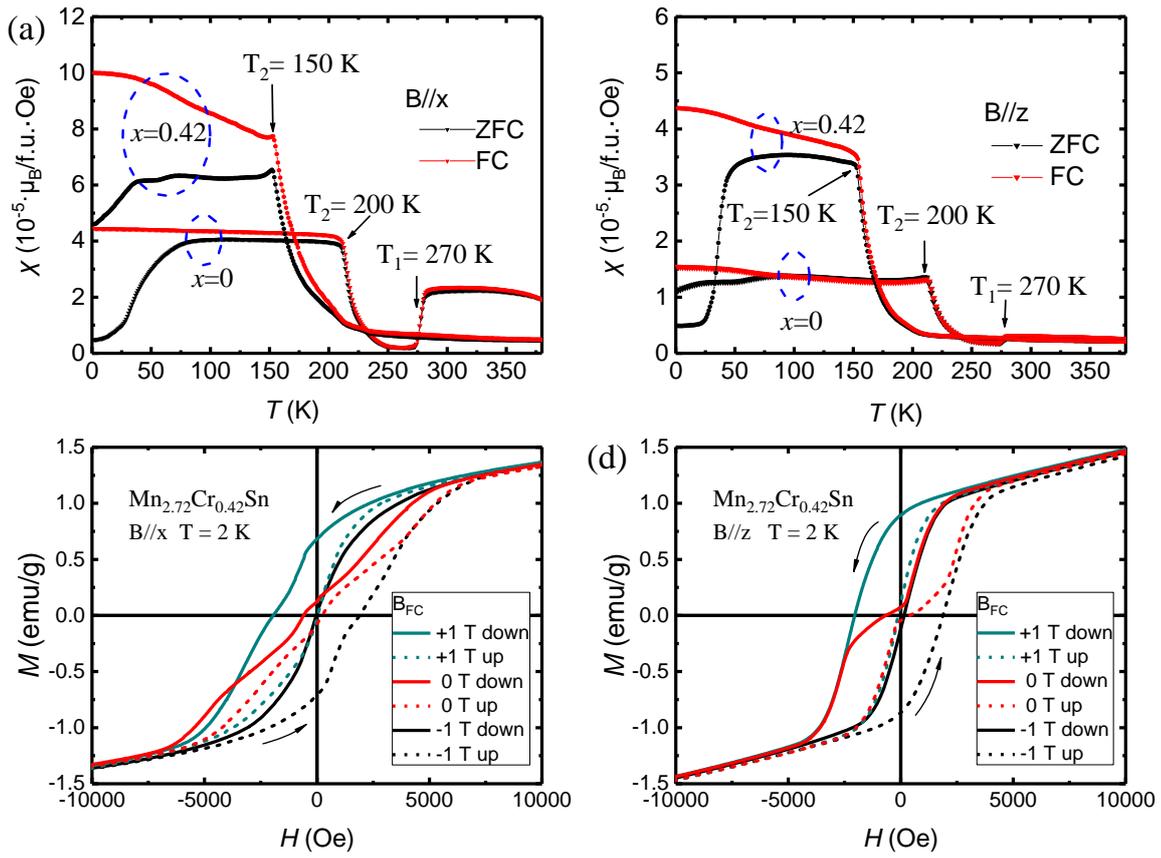

Fig.3

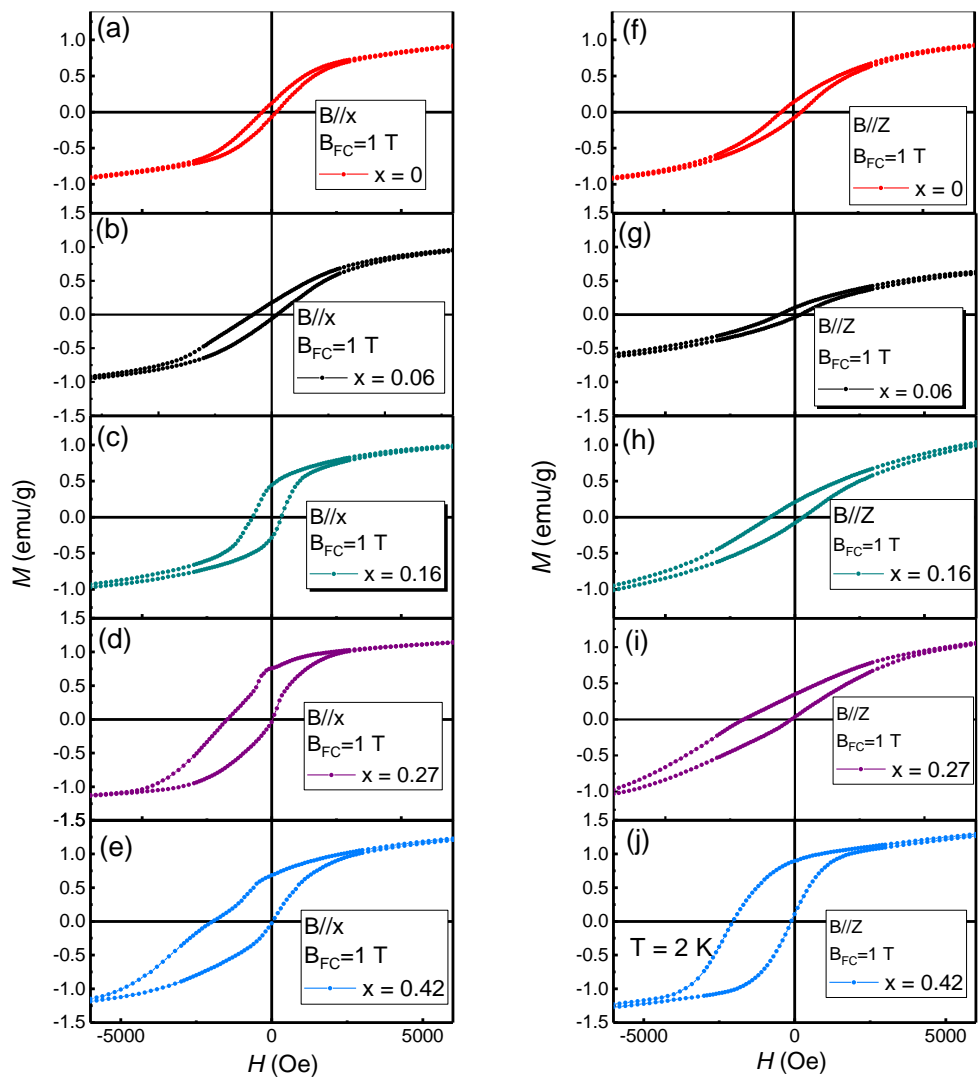

Fig.4.

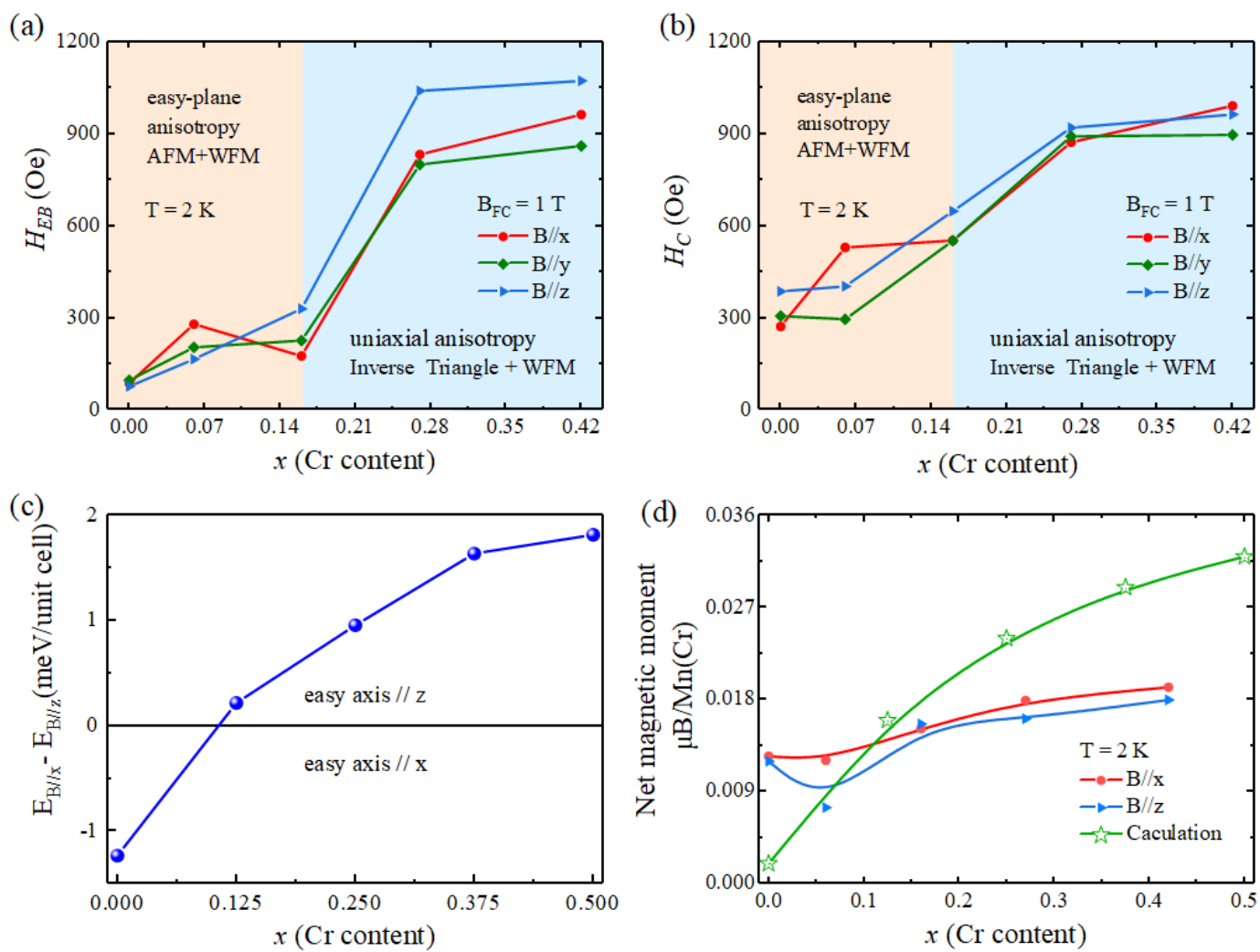

Fig.5

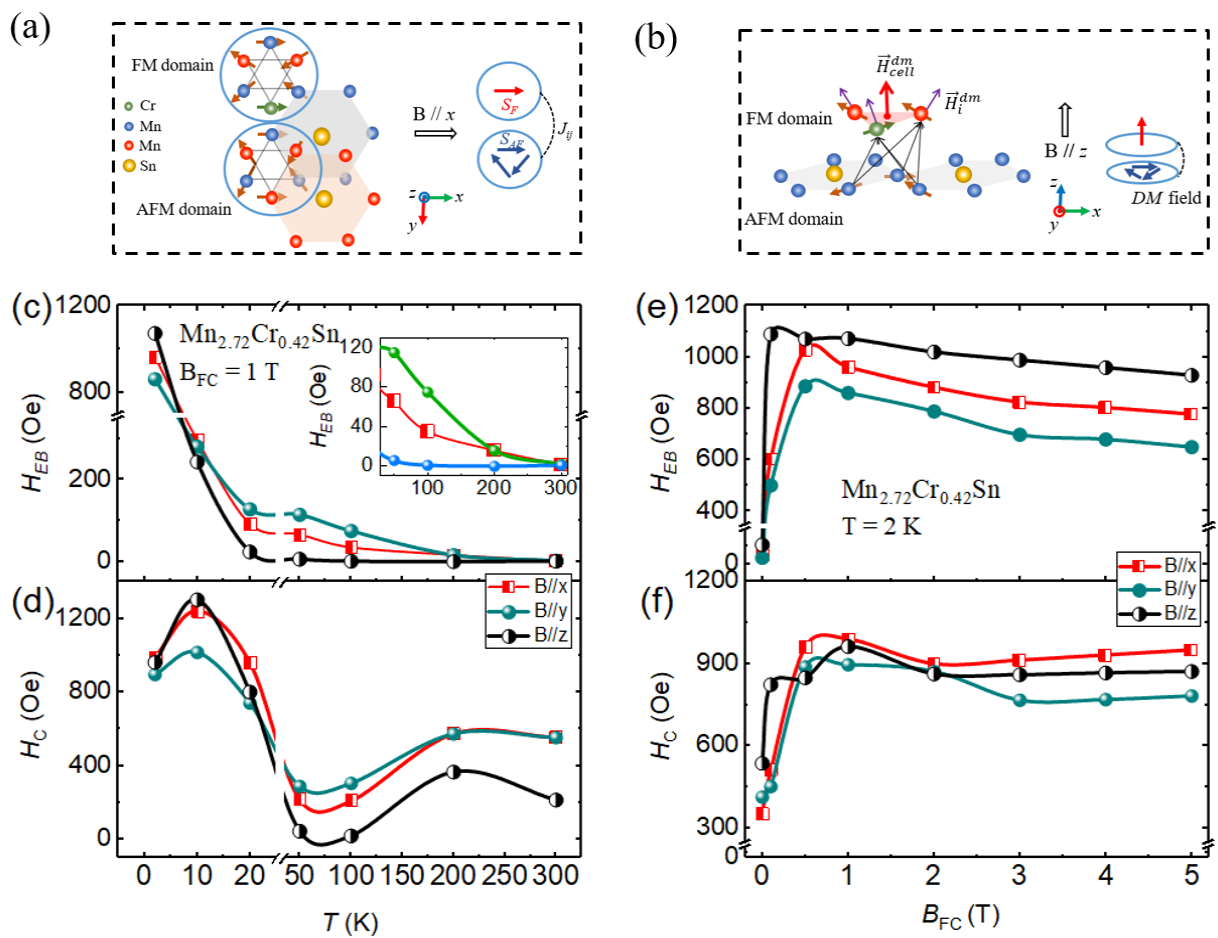

# Supplementary materials

**Coexistence of in-plane and out-of-plane exchange Bias in correlated kagome antiferromagnet $Mn_{3-x}Cr_xSn$**

Xiaoyan Yang[1], Yi Qiu[2], Jie Su[2], Jincang Zhang[1, 3, †], Xiao Liang[4, ‡], Guixin Cao[1, *]

[1] Materials Genome Institute, Shanghai University, 200444 Shanghai, China

[2] College of Chemistry and Molecular Engineering, Peking University, Beijing 100871, China

[3] Zhejiang Laboratory, Hangzhou 311100, China

[3] Information Materials and Device Applications Key Laboratory of Sichuan Provincil Universities, Chengdu University of Information Technology, 610225 Sichuan, China

[*] Corresponding author, G. Cao, Email: guixincao@shu.edu.cn
[‡] Corresponding author, X. Liang, E-mail: liangxiao920@163.com
[†] Corresponding author, J. Zhang, E-mail: jczhang@shu.edu.cn

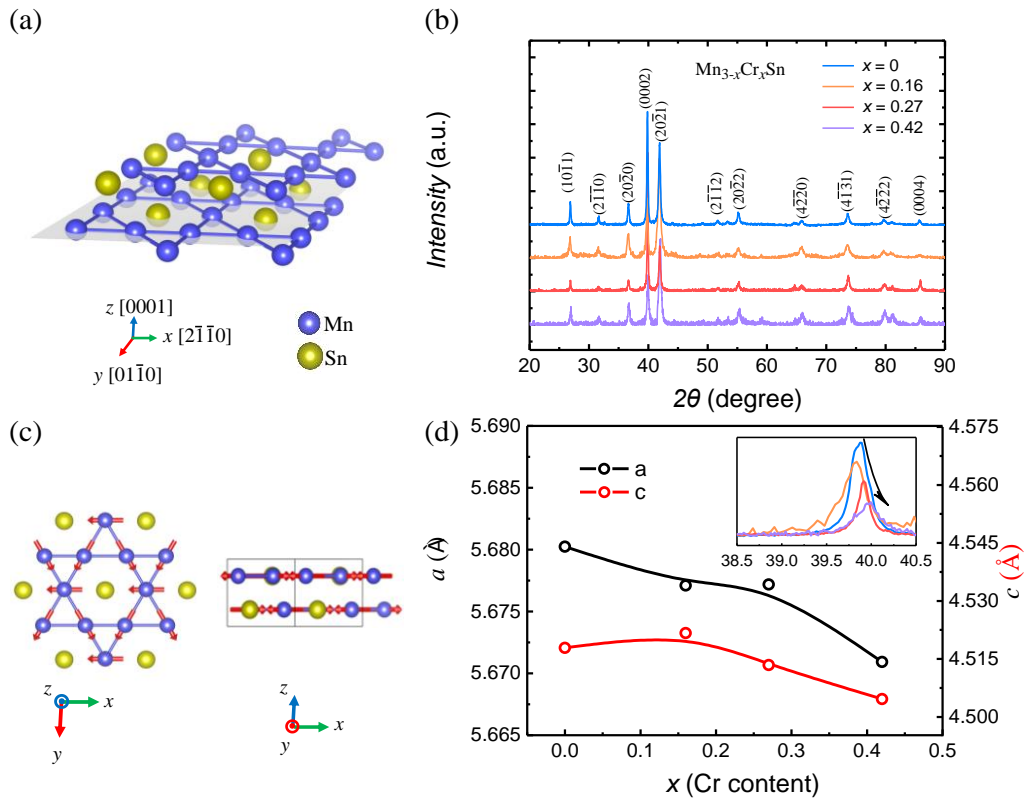

Fig.1S: Magnetic texture in the kagome lattice. Here, we take the $x$, $y$ and $z$ coordinates along $[2\bar{1}10]$, $[01\bar{1}0]$ and $[0001]$. Mn moments (~3 μB, μB is the Bohr magneton) lying in each $x$–$y$ kagome plane form a 120° degree structure. Crystal structure of Hexagonal $Mn_3Sn$, blue and yellow spheres represent Mn and Sn atoms. Sn sits at the center of the Mn hexagon ring in a-b plane stacking along the c axis. (b) Inverse triangular spin structure of $Mn_3Sn$. (c) The powder XRD patterns of $Mn_{3-x}Cr_xSn$ single crystals at 300 K. (d) The Cr content dependence of lattice parameters a and c, the (0001) peak of different Cr-doping crystals is shown in the inset.

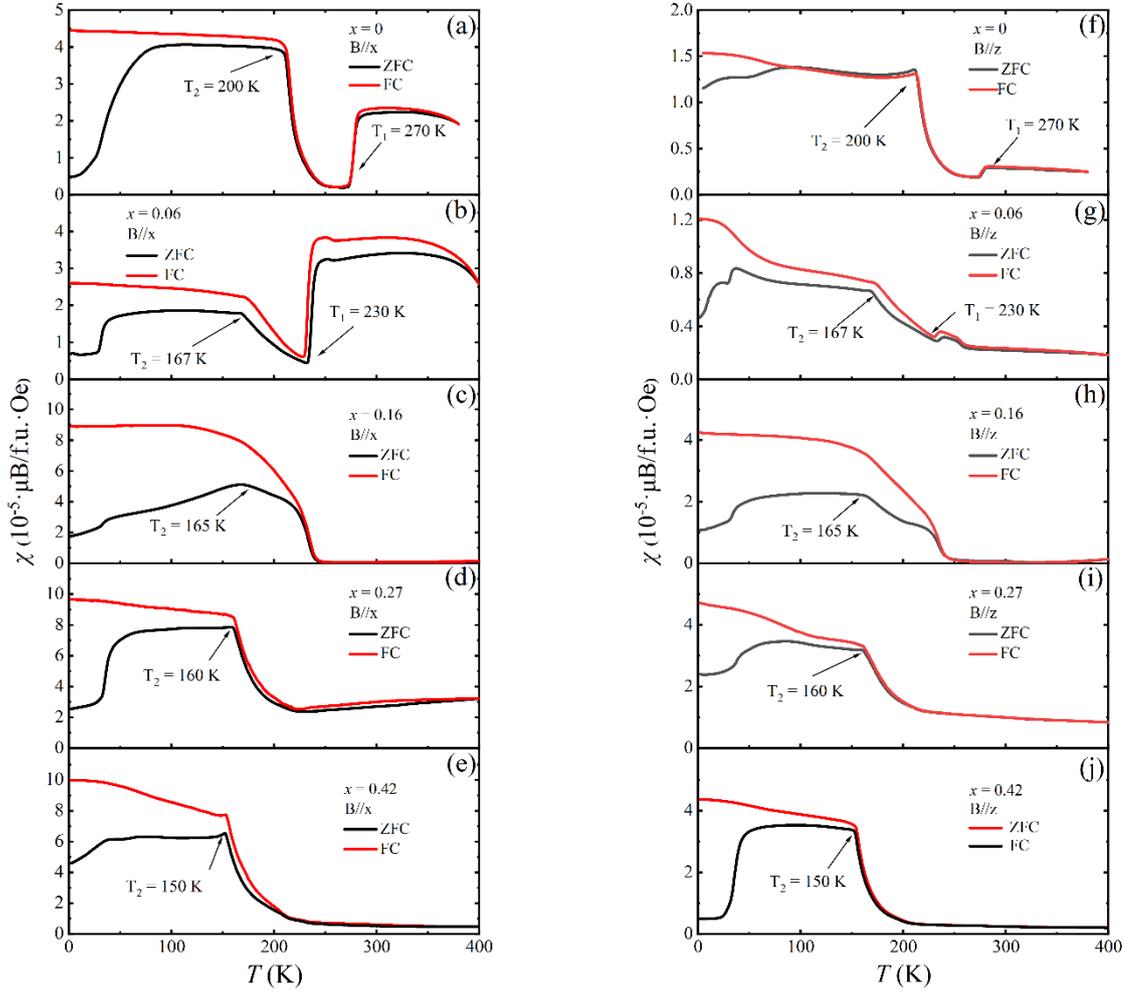

Fig 2S. Temperature dependence of magnetization $M(T)$ measured in zero field cooled (ZFC, black lines) and field cooled (FC, red lines) modes for $Mn_{3-x}Cr_xSn$ ($x$ = 0, 0.06, 0.16, 0.27, 0.42). (a-e) The field is parallel to $x$ direction; (f-j) The field is parallel to $z$ direction.

$Mn_3Sn$ undergoes an inverse triangular AFM to spiral phase transition at $T_1$ = 270 K as exemplified in the magnetization $M(T)$ in Fig. 1(a). With temperature decreasing further, $M(T)$ exhibits a ferromagnetic-like transition at $T_2$ = 200 K. The phase transition temperature $T_1$ and ferromagnetic-like transition $T_2$ decrease with Cr content increasing. Then, the spiral state transition in $Mn_{3-x}Cr_xSn$ is suppressed with Cr content $x$ = 0.16, as shown in Fig. 1(c). Compared with pure phase, the Cr-doped $Mn_{3-x}Cr_xSn$ spiral state has disappeared at low temperature, producing an inverse triangular spin structure.

Table IS: The calculated spin structure and net moment for $Mn_3Sn$.

|  | Mn1 | Mn2 | Mn3 | Mn4 | Mn5 | Mn6 |
| --- | --- | --- | --- | --- | --- | --- |
| $m_x$ ($\mu_B$) | 3.167 | -1.586 | -1.579 | 3.166 | -1.583 | -1.582 |
| $m_y$ ($\mu_B$) | 0.003 | -2.741 | 2.745 | 0.000 | -2.743 | 2.743 |
| $m_z$ ($\mu_B$) | -0.001 | -0.003 | -0.003 | 0.000 | -0.002 | -0.002 |
| $\alpha$ ($\mu_B$) | 0.057 | 120.055 | 119.909 | 0.000 | 119.989 | 119.974 |
| $\beta$ ($\mu_B$) | 89.946 | 149.945 | 29.909 | 90.000 | 150.011 | 29.974 |
| $\gamma$ ($\mu_B$) | 90.018 | 90.054 | 90.054 | 90.000 | 90.036 | 90.036 |

Net magnetic moment: 0.002 $\mu_B$/Mn

Table IIS: The calculated spin structure and net moment for $Mn_{3-x}Cr_xSn$ with $x=0.125$.

|                | Cr1 | Mn2 | Mn3 | Mn4 | Mn5 | Mn6 |
|----------------|-----|-----|-----|-----|-----|-----|
| $m_x$ ($\mu_B$) | 3.040 | -1.632 | -1.636 | 3.175 | -1.711 | -1.715 |
| $m_y$ ($\mu_B$) | 0.012 | -2.525 | 2.527 | 0.005 | -2.612 | 2.615 |
| $m_z$ ($\mu_B$) | 0.053 | -0.021 | -0.021 | 0.017 | -0.022 | -0.018 |
| $\alpha$ ($\mu_B$) | 1.024 | 122.875 | 122.918 | 0.320 | 123.226 | 123.258 |
| $\beta$ ($\mu_B$) | 89.774 | 147.122 | 32.922 | 89.910 | 146.771 | 33.260 |
| $\gamma$ ($\mu_B$) | 89.001 | 90.400 | 90.400 | 89.693 | 90.404 | 90.330 |

Net magnetic moment: 0.016 $\mu_B$/Mn

With Cr substitution content increasing to $x=0.125$, 0.25, 0.375 and 0.5, the net magnetic moments increase to 0.016 uB, 0.024 uB, 0.029 uB and 0.032uB per Mn (Cr) atom, which is agree with our measurements of saturation magnetization in Cr doped $Mn_3Sn$ samples, resulting in the magnetic exchange effect increased. On the other hand, taking $Cr_{0.125}Mn_{2.875}Sn$ as an example, the calculated magnetic moment of Cr atom is 3.04 uB, which is a little lower than Mn atoms, and the spin directions of Mn nearby the Cr atoms tilted both in in-plane and out-plane directions. To facilitate comparison, the calculations results are shown with the green pentagram in Fig.4 (d).